\documentclass{article}

\usepackage{graphicx} % For including figures
\usepackage{braket}
\usepackage{physics}
\usepackage{tikz}
\usepackage{tikz-3dplot}
\usepackage{float}
\usepackage{placeins}
\usepackage{url}

\begin{document}

\sloppy

\title{Entanglement on Two Bloch Spheres: \\Exploring Two-Qubit Stabilizer Group Structure.}

\author{Stanislav Filatov and Marcis Auzinsh}
\maketitle
\begin{abstract}
    In this paper, we explore the graphical representation of two-qubit entanglement on two Bloch Spheres via stabilizer formalism. We relate the density matrix to the graphical representation on two Bloch Spheres by showing how both may be derived from the stabilizer group structure of the state. Then we use the representation to explore the symmetries present in maximally entangled states.
\end{abstract}


\section{Introduction}
Bloch Sphere representation \cite{Bloch1946, Feynman1957, Radcliffe1971, Arecchi1972} is a powerful visualization tool that gets used across various areas of Quantum Mechanics. Extension of this representation to systems of two or more qubits is challenging due to entanglement. 

There is a number of works that focus on different aspects of two-qubit system graphical representation, the need to project higher-dimensional space creates multitude of possibilities: from sections of three-dimensional space  \cite{Bengtsson2006, Morelli2024} to examination of fibrations \cite{Mosseri2001, Bernevig2003, Mosseri2006, Uskov2008} to geometric algebra \cite{Rau2021, Wie2020}. Among scalable approaches Majorana stellar representation should also be noted  \cite{Majorana1932, Aulbach2010} that allows to represent N-level system by N-1 vectors on the Bloch Sphere resulting in versatility and variety of applications \cite{Cormann2017, Yu-Guo2021, Xingyu2021}. 

The authors would like to contribute to the aforementioned labour by proposing a graphical representation where any pure two-qubit state may be represented on two Bloch Spheres. This method assigns a Bloch Sphere to each qubit, even in the case of maximally entangled states. As a result, local rotations on a single qubit of an entangled pair still correspond to rotations of a single Bloch Sphere.

In the previous work \cite{SFMA2024} we have been exploring the two-qubit system through the lens of Geometric Algebra. Now we focus on the Stabilizer group structure and show how it contains the method to represent two-qubit entanglement on two Bloch Spheres. This stabilizer-based framework reveals a straightforward way to connect density matrix of a state with its graphical representation on two Bloch Spheres and demonstrates how Two Bloch Sphere approach is reflecting the symmetries of two-partite Hilbert space. Furthermore, grounding the representations in stabilizer formalism makes them accessible to the quantum computation and quantum foundations communities, where intuitive tools for qubit manipulation are most needed.

The work is structured as follows. Chapter 2 introduces stabilizer states and stabilizer groups. In Chapter 3, we explore the connection between the stabilizer group of a state and its density matrix. Chapter 4, examines the relationship between the stabilizer group and its graphical representation, ultimately establishing the link between a state's density matrix and its representation on two Bloch Spheres. Chapter 5 discusses the underlying principles of this correspondence and provides a use case for the method. Finally, Chapter 6 concludes the work. We emphasize examples and visualizations, allowing readers to engage with graphical representations and cultivate visual intuition as they progress through the paper.

\section{Two-Qubit Stabilizer States and Groups}
This chapter presents stabilizer states and groups to ensure the paper remains self-contained. Rather than providing an exhaustive treatment, we focus on a concise overview that aligns with our approach and facilitates understanding of the concepts discussed later in the work. For a more rigorous account of the topic, we direct readers to \cite{NielsenChuang2011} Chapter 10 or \cite{Gottesman1997}.

Essentially, two-qubit stabilizer states are those states that are reachable from the state $\ket{\uparrow\uparrow}$ by Clifford gates. Clifford gates are a specific set of quantum gates that map the Pauli group to itself under conjugation, thereby preserving the commutation relations between Pauli operators. This class includes gates that can be expressed as unitary matrices satisfying the equation $U P U^\dagger \in \text{Pauli}$ for any Pauli operator $P$. A representative set of Clifford gates is comprised of the Hadamard gate, the Phase gate, and the CNOT gate. 

To illustrate this concept further, consider the set of $4\times4$ two-qubit Pauli matrices that have a form $ I \otimes P$ or $ P \otimes I$ or $P \otimes P$ where $I$ is $2 \times 2$ Identity and P is some $2\times2$ Pauli matrix $X, Y, Z$. Now let us navigate the space of pure two-qubit states by repeatedly applying these matrices to the state $\ket{\uparrow\uparrow}$ and restricting the angle of rotation to $90$ degrees. The states reachable in this way are the two-qubit stabilizer states. There are 60 two-qubit stabilizer states: 36 separable and 24 maximally entangled ones \cite{Gross2006, Garcia2014}. 

Stabilizer group of a given state is formed by matrices that, when applied to the given state, leave it unchanged. Essentially the members of the stabilizer group, besides Identity, are eigenmatrices of a given state times the eigenvalue. Below are correspondences between some of the stabilizer states and their stabilizer groups. We have chosen four states, two separable and two entangled: $\ket{\uparrow\uparrow}$; $\ket{\uparrow\leftarrow} = \frac{1}{\sqrt{2}}(\ket{\uparrow\uparrow} - \ket{\uparrow\downarrow})$; $\ket{\Psi^-} = \frac{1}{\sqrt{2}}(\ket{\uparrow\downarrow} - \ket{\downarrow\uparrow})$ and $\ket{\Phi^{+i}} = \frac{1}{\sqrt{2}}(\ket{\uparrow\uparrow} + i\ket{\downarrow\downarrow})$

\begin{equation}
\begin{aligned}
    \ket{\uparrow\uparrow} \leftrightarrow \ &\{I \otimes I; Z \otimes I; I \otimes Z; Z \otimes Z\} \\
    \ket{\uparrow\leftarrow} \leftrightarrow \  &\{I \otimes I; Z \otimes I; - I \otimes X; - Z \otimes X\} \\
    \ket{\Psi^-} \leftrightarrow \  &\{I \otimes I; - Z \otimes Z; -X \otimes X; -Y \otimes Y\}\\
    \ket{\Phi^{+i}} \leftrightarrow \  &\{I \otimes I; Z \otimes Z; X \otimes Y; Y \otimes X\}\\
\end{aligned}
\label{eq:state-group}
\end{equation}

The stabilizer group associated with a pure two-qubit state consists of four matrices, one of which is always the \(4 \times 4\) identity matrix. Separable states include one matrix of the type \(P \otimes P\), one of the type \(P \otimes I\), and one of the type \(I \otimes P\). Conversely, the stabilizer group of an entangled state contains the identity matrix and three matrices of the type \(P \otimes P\).

\section{Connecting stabilizer group and density matrix}
The state's density matrix may be related to the stabilizer group of a state in a straightforward way. One just needs to sum the matrices that make up the stabilizer group of a state. More formally, given n-qubit state $\ket{\Psi}$ and its stabilizer group $\mathcal{S}$ consisting of $2^n$ matrices $M \in \mathcal{S}$ s.t. $M\ket{\Psi} = \ket{\Psi}$ for all $M$ in $\mathcal{S}$ the following is true \cite{Wu2015}:

\begin{equation}
    \begin{aligned}
        \rho_{\mathcal{S}} = \ket{\Psi}\bra{\Psi} = \frac{1}{2^n}\sum\limits_{M \in \mathcal{S}}M
    \end{aligned}
\label{eq:rho-group-sum}    
\end{equation}

With the density matrix expressed in terms of the stabilizer group, we can now explore how the stabilizer group connects to the graphical representation of the state.

\FloatBarrier

\section{Connecting stabilizer group to graphical representation}
Let us examine the stabilizer groups of the states from Equation \ref{eq:state-group} and interpret them in terms of graphical representations on two Bloch Spheres. We will start with examination of separable states first where the graphical interpretation is more straightforward as it involves two Bloch Sphere representations of single qubits. We then use insights obtained from examination of separable states to devise a way to represent maximally entangled states. 

We focus on two-qubit Stabilizer states in this chapter and generally in the work as this set of states is rich enough to contain maximally entangled states and at the same time simple enough to provide straightforward interpretations of the stabilizer group structure. The extension of our method is discussed at the end of the chapter. 

\subsection{Separable states}
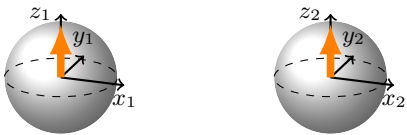
\begin{figure}[H]
  \centering
\begin{minipage}[b][2cm][t]{0.2\textwidth}
\centering
\tdplotsetmaincoords{70}{20}
\begin{tikzpicture}[tdplot_main_coords, scale=0.75]
    % Draw the Bloch sphere with reduced opacity
    \shade[ball color=white!50, opacity=0.6] (0,0,0) circle (1cm);
    
    % Draw axes with adjusted scaling
    \draw[thick,->, scale=1] (0,0,0) -- (1.2,0,0) node[anchor=north, font=\small]{$x_1$};
    \draw[thick,->, scale=1] (0,0,0) -- (0,1.2,0) node[anchor=south, font=\small]{$y_1$};
    \draw[thick,->, scale=1] (0,0,0) -- (0,0,1.2) node[anchor=east, font=\small]{$z_1$};
    
    % Draw equator (dotted line) on the back side with adjusted scaling
    %\tdplotdrawarc[thick, scale=1]{(0,0,0)}{1}{-90}{90}{}{}
    
    % Draw equator (full line) on the front side with adjusted scaling
    \tdplotdrawarc[thin, dashed, scale=1]{(0,0,0)}{1}{-4}{356}{}{}
    
        % Define the arbitrary state vector coordinates
    \coordinate (statevector) at (0, 0, 1);
    
    % Draw the state vector as a thicker arrow
    \draw[orange, line width=1mm, -latex] (0,0,0) -- (statevector);
\end{tikzpicture}
\end{minipage}%
\begin{minipage}[b][2cm][t]{0.2\textwidth}
\centering
\tdplotsetmaincoords{70}{20}
\begin{tikzpicture}[tdplot_main_coords, scale=0.75]
    % Draw the Bloch sphere with reduced opacity
    \shade[ball color=white!50, opacity=0.6] (0,0,0) circle (1cm);
    
    % Draw axes with inverted directions
    \draw[thick,->, scale=1] (0,0,0) -- (1.2,0,0) node[anchor=north, font=\small]{$x_2$};
    \draw[thick,->, scale=1] (0,0,0) -- (0,1.2,0) node[anchor=south, font=\small]{$y_2$};
    \draw[thick,->, scale=1] (0,0,0) -- (0,0,1.2) node[anchor=east, font=\small]{$z_2$};
    
    % Draw equator (dotted line) on the back side with adjusted scaling
    %\tdplotdrawarc[thick, scale=1]{(0,0,0)}{1}{-90}{90}{}{}
    
    % Draw equator (full line) on the front side with adjusted scaling
    \tdplotdrawarc[thin, dashed, scale=1]{(0,0,0)}{1}{-4}{356}{}{}
    
        % Define the arbitrary state vector coordinates
    \coordinate (statevector) at (0, 0, 1);
    
    % Draw the state vector as a thicker arrow
    \draw[orange, line width=1mm, -latex] (0,0,0) -- (statevector);
\end{tikzpicture}
\end{minipage}

  \caption{Two Bloch Sphere representation of the separable state $\ket{\uparrow\uparrow}$}
  \label{fig:uu}
\end{figure}

The stabilizer group of the state $\ket{\uparrow\uparrow}$ is $\{I \otimes I; Z \otimes I; I \otimes Z; Z \otimes Z\}$. The representation of this state may be seen in Figure \ref{fig:uu}. Let us now tie the graphical representation and the stabilizer group structure.  We propose to interpret structure like $Z \otimes I$ as indicating which of the first Bloch Sphere axes is aligned/anti-aligned with the statevector arrow. In this case $z_1$ axis is aligned with the statevector arrow. In the same way we arrive at the interpretation of $I \otimes Z$: $z_2$ axis is aligned with the statevector arrow. Furthermore, $Z \otimes Z$ then can be interpreted in terms of mutual co-alignment of $z_1$ and $z_2$ axes. Note that just like in the group structure $Z \otimes Z$ is implied by generators $Z \otimes I$ and  $I \otimes Z$, the mutual coalignment of axes is implied by the first two interpretations. 

Let us now create graphical representation of the state $\ket{\uparrow\leftarrow}$ with stabilizer group $\{I \otimes I; Z \otimes I; - I \otimes X; - Z \otimes X\}$. $Z \otimes I$ implies, just as in the case above that the axis $z_1$ will face along the statevector arrow; $- I \otimes X$ implies that the axis $x_2$ will be anti-aligned with the statevector arrow; $- Z \otimes X$ implies that the axis $z_1$ will be anti-aligned with the axis $x_2$ or, that $z_1$ will be facing in the same direction as $-x_2$. Figure \ref{fig:ul} contains the representation. Indeed that is a representation of the state $\ket{\uparrow\leftarrow}$. 

We have chosen the statevector arrows to remain static, therefore we define interpretations in terms of coordinate axes. In fact, such representation allows for the most straightforward interpretation of the stabilizer group structure because we effectively have only one direction of the statevector arrows. The rotations of single qubits are represented as rotations of the coordinate axes, i.e. passive rotations. There are $6$: $\pm x_1, \pm y_1, \pm z_1$ labels that may face in the direction of the statevector on the first Bloch Sphere and $6$ on the second. As a result there are $6 \times 6 = 36$ configurations which correspond to the number of separable stabilizer states.

\begin{figure}[H]
  \centering
\begin{minipage}[b][2cm][t]{0.2\textwidth}
\centering
\tdplotsetmaincoords{70}{20}
\begin{tikzpicture}[tdplot_main_coords, scale=0.75]
    % Draw the Bloch sphere with reduced opacity
    \shade[ball color=white!50, opacity=0.6] (0,0,0) circle (1cm);
    
    % Draw axes with adjusted scaling
    \draw[thick,->, scale=1] (0,0,0) -- (1.2,0,0) node[anchor=north, font=\small]{$x_1$};
    \draw[thick,->, scale=1] (0,0,0) -- (0,1.2,0) node[anchor=south, font=\small]{$y_1$};
    \draw[thick,->, scale=1] (0,0,0) -- (0,0,1.2) node[anchor=east, font=\small]{$z_1$};
    
    % Draw equator (dotted line) on the back side with adjusted scaling
    %\tdplotdrawarc[thick, scale=1]{(0,0,0)}{1}{-90}{90}{}{}
    
    % Draw equator (full line) on the front side with adjusted scaling
    \tdplotdrawarc[thin, dashed, scale=1]{(0,0,0)}{1}{-4}{356}{}{}
    
        % Define the arbitrary state vector coordinates
    \coordinate (statevector) at (0, 0, 1);
    
    % Draw the state vector as a thicker arrow
    \draw[orange, line width=1mm, -latex] (0,0,0) -- (statevector);
\end{tikzpicture}
\end{minipage}%
\begin{minipage}[b][1.75cm][t]{0.2\textwidth}
\centering
\tdplotsetmaincoords{70}{20}
\begin{tikzpicture}[tdplot_main_coords, scale=0.75]
    % Draw the Bloch sphere with reduced opacity
    \shade[ball color=white!50, opacity=0.6] (0,0,0) circle (1cm);
    
    % Draw equator (full line) on the front side with adjusted scaling
    \tdplotdrawarc[thin, dashed, scale=1]{(0,0,0)}{1}{-4}{356}{}{}    
    
    % Draw axes with inverted directions
    \draw[thick,->, scale=1] (0,0,0) -- (1.2,0,0) node[anchor=north, font=\small]{$z_2$};
    \draw[thick,->, scale=1] (0,0,0) -- (0,1.2,0) node[anchor=south, font=\small]{$y_2$};
    \draw[thick,->, scale=1] (0,0,0) -- (0,0,-1.2) node[anchor=west, font=\small]{$x_2$};
    
    % Draw equator (dotted line) on the back side with adjusted scaling
    %\tdplotdrawarc[thick, scale=1]{(0,0,0)}{1}{-90}{90}{}{}

        % Define the arbitrary state vector coordinates
    \coordinate (statevector) at (0, 0, 1);
    
    % Draw the state vector as a thicker arrow
    \draw[orange, line width=1mm, -latex] (0,0,0) -- (statevector);
\end{tikzpicture}
\end{minipage}

  \caption{Two Bloch Sphere representation of the separable state $\ket{\uparrow\leftarrow}$}
  \label{fig:ul}
\end{figure}

\subsection{Maximally entangled states}

 Now, let’s focus on maximally entangled stabilizer states. The stabilizer group for these states consists only of matrices of the form $P \otimes P$. We can apply the interpretation we established for separable states to understand these structures. Specifically, this will help us determine which axes of the two Bloch Spheres are aligned.

The stabilizer group of the state $\ket{\Psi^-}$ is $\{I \otimes I; - Z \otimes Z; -X \otimes X; -Y \otimes Y\}$. Interpreting $- Z \otimes Z$ allows us to conclude that in a graphical representation $z_1$ is facing in the same direction as $-z_2$. In other words, $z_1$ and $z_2$ axes are anti-aligned. Interpreting the other members of the group we arrive at a representation in which all the coordinate axes of the second BS are anti-aligned with the coordinate axes of the first BS (Figure \ref{fig:Psi-}). Having inverted odd number $(3)$ of coordinate axes on the second BS we obtain left-handed coordinates on it. 

\begin{figure}[H]
  \centering
\begin{minipage}[b][2cm][t]{0.2\textwidth}
\centering
\tdplotsetmaincoords{70}{20}
\begin{tikzpicture}[tdplot_main_coords, scale=0.75]
    % Draw the Bloch sphere with reduced opacity
    \shade[ball color=white!50, opacity=0.6] (0,0,0) circle (1cm);
    
    % Draw axes with adjusted scaling
    \draw[thick,->, scale=1] (0,0,0) -- (1.2,0,0) node[anchor=north, font=\small]{$x_1$};
    \draw[thick,->, scale=1] (0,0,0) -- (0,1.2,0) node[anchor=south, font=\small]{$y_1$};
    \draw[thick,->, scale=1] (0,0,0) -- (0,0,1.2) node[anchor=east, font=\small]{$z_1$};
    
    % Draw equator (dotted line) on the back side with adjusted scaling
    %\tdplotdrawarc[thick, scale=1]{(0,0,0)}{1}{-90}{90}{}{}
    
    % Draw equator (full line) on the front side with adjusted scaling
    \tdplotdrawarc[thin, dashed, scale=1]{(0,0,0)}{1}{-4}{356}{}{}
    
    % Draw the state vector as a thicker arrow with adjusted scaling
    \shade[ball color=orange, opacity=1, scale=2] (0,0,0) circle (0.05cm);
\end{tikzpicture}
\end{minipage}%
\begin{minipage}[b][1.75cm][t]{0.2\textwidth}
\centering
\tdplotsetmaincoords{70}{20}
\begin{tikzpicture}[tdplot_main_coords, scale=0.75]
    % Draw the Bloch sphere with reduced opacity
    \shade[ball color=white!50, opacity=0.6] (0,0,0) circle (1cm);
    
    % Draw axes with inverted directions
    \draw[thick,->, scale=1] (0,0,0) -- (-1.2,0,0) node[anchor=north, font=\small]{$x_2$};
    \draw[thick,->, scale=1] (0,0,0) -- (0,-1.2,0) node[anchor=north, font=\small]{$y_2$};
    \draw[thick,->, scale=1] (0,0,0) -- (0,0,-1.2) node[anchor=west, font=\small]{$z_2$};
    
    % Draw equator (dotted line) on the back side with adjusted scaling
    %\tdplotdrawarc[thick, scale=1]{(0,0,0)}{1}{-90}{90}{}{}
    
    % Draw equator (full line) on the front side with adjusted scaling
    \tdplotdrawarc[thin, dashed, scale=1]{(0,0,0)}{1}{-4}{356}{}{}
    
    % Draw the state vector as a thicker arrow with adjusted scaling
    \shade[ball color=orange, opacity=1, scale=2] (0,0,0) circle (0.05cm);
\end{tikzpicture}
\end{minipage}

  \caption{Two Bloch Sphere representation of the entangled state $\ket{\Psi^-}$}
  \label{fig:Psi-}
\end{figure}

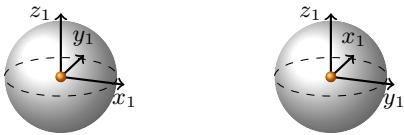
\begin{figure}[H]
  \centering
\begin{minipage}[b][2cm][t]{0.2\textwidth}
\centering
\tdplotsetmaincoords{70}{20}
\begin{tikzpicture}[tdplot_main_coords, scale=0.75]
    % Draw the Bloch sphere with reduced opacity
    \shade[ball color=white!50, opacity=0.6] (0,0,0) circle (1cm);
    
    % Draw axes with adjusted scaling
    \draw[thick,->, scale=1] (0,0,0) -- (1.2,0,0) node[anchor=north, font=\small]{$x_1$};
    \draw[thick,->, scale=1] (0,0,0) -- (0,1.2,0) node[anchor=south, font=\small]{$y_1$};
    \draw[thick,->, scale=1] (0,0,0) -- (0,0,1.2) node[anchor=east, font=\small]{$z_1$};
    
    % Draw equator (dotted line) on the back side with adjusted scaling
    %\tdplotdrawarc[thick, scale=1]{(0,0,0)}{1}{-90}{90}{}{}
    
    % Draw equator (full line) on the front side with adjusted scaling
    \tdplotdrawarc[thin, dashed, scale=1]{(0,0,0)}{1}{-4}{356}{}{}
    
    % Draw the state vector as a thicker arrow with adjusted scaling
    \shade[ball color=orange, opacity=1, scale=2] (0,0,0) circle (0.05cm);
\end{tikzpicture}
\end{minipage}%
\begin{minipage}[b][2cm][t]{0.2\textwidth}
\centering
\tdplotsetmaincoords{70}{20}
\begin{tikzpicture}[tdplot_main_coords, scale=0.75]
    % Draw the Bloch sphere with reduced opacity
    \shade[ball color=white!50, opacity=0.6] (0,0,0) circle (1cm);
    
    % Draw axes with inverted directions
    \draw[thick,->, scale=1] (0,0,0) -- (1.2,0,0) node[anchor=north, font=\small]{$y_1$};
    \draw[thick,->, scale=1] (0,0,0) -- (0,1.2,0) node[anchor=south, font=\small]{$x_1$};
    \draw[thick,->, scale=1] (0,0,0) -- (0,0,1.2) node[anchor=east, font=\small]{$z_1$};
    
    % Draw equator (dotted line) on the back side with adjusted scaling
    %\tdplotdrawarc[thick, scale=1]{(0,0,0)}{1}{-90}{90}{}{}
    
    % Draw equator (full line) on the front side with adjusted scaling
    \tdplotdrawarc[thin, dashed, scale=1]{(0,0,0)}{1}{-4}{356}{}{}
    
    % Draw the state vector as a thicker arrow with adjusted scaling
    \shade[ball color=orange, opacity=1, scale=2] (0,0,0) circle (0.05cm);
\end{tikzpicture}
\end{minipage}

  \caption{Two Bloch Sphere representation of the entangled state $\ket{\Phi^{+i}}$}
  \label{fig:Phi+i}
\end{figure}

The stabilizer group for the state $\ket{\Phi^{+i}}$ is $\{I \otimes I; Z \otimes Z; X \otimes Y; Y \otimes X\}$. The group structure consisting only of the members of the form $P \otimes P$ indicates that the state is maximally entangled. Interpretation of the three members of the group gives us the following pairs of co-directional axes: $\{z_1 \And z_2$; $x_1 \And y_2$; $y_1 \And x_2\}$ resulting in the representation shown in Figure \ref{fig:Phi+i}. Note that the second BS has left-handed coordinates.

These representations do not have statevector arrows at all, or rather they are facing in the $(0,0,0)$ direction. This is consistent with the fact that reduced density matrix of a single qubit in a maximally entangled state is Identity. As a result, the representation of the maximally entangled state actually requires less structure than the representation of a separable state. All the information about the entangled state is encoded in the \textit{relative} directions of the coordinate axes. Another important feature of the maximally entangled state representation is the different handedness of Bloch Sphere coordinates. 

Representing entanglement through the \textit{relative} alignment of coordiante axes of the two Bloch Spheres makes it obvious not only why the state of single qubit of the entangled pair is maximally mixed, but also why we need to have access to the whole system to extract any information about the two-qubit state. Looking at a single Bloch Sphere in Figures \ref{fig:Psi-} and \ref{fig:Phi+i} does not reveal anything about the relative alignment of the cordinate axes and therefore about the state.

Moreover, this representation is consistent with the convention we have chosen before of representing local rotations through passive rotations rather than through statevector rotations. They work even in the case of maximally entangled state when the statevector arrows are dots in the middle of BS and allow to keep track of the evolution of the state of the system. 

Although the local rotations refer to coordinate rotations on a single BS, if we start from a maximally entangled state and restrict ourselves to only local rotations of a single qubit - we can still reach any \textit{relative} alignment of the coordinate axes and therefore any maximally entangled state. The representations in Figures \ref{fig:Psi-} and \ref{fig:Phi+i} clarify this quantum mechanical fact. In Chapter 6 we come back to this idea to investigate symmetries of entangled states using local rotations and stabilizer formalism.

Let us count how many maximally entangled two-qubit stabilizer states result from such representation convention. Having fixed the coordinates of the first BS, the $z_2$ axis may face any of the 6 directions; for each of those $x_2$ axis may face 4 possible directions because it has to also be orthogonal to $z_2$; the combination of $z_2$ and $x_2$, taking into account the handedness, defines the direction of $y_2$ unambiguously. Therefore there are $6 \times 4 = 24$ possible configurations, which corresponds to the number of maximally entangled two-qubit stabilizer states.

\subsection{Summary}

In this way, the representation of all $60 = 36 + 24$ stabilizer states can be obtained. While not the focus of this work, the method for representing stabilizer states can be extended to any arbitrary pure two-qubit state. The ``stabilizer" group of a non-stabilizer state can be expressed as a sum of appropriately scaled stabilizer groups of stabilizer states. The graphical representation is then achieved by scaling and combining the representations of the corresponding stabilizer states. 

Linking stabilizer group to graphical representation on one hand and to density matrix on the other - we effectively link the latter two. In the next chapter we will speak about rotations to explain how the movement between the states happens, show some useful applications of the two BS formalism and reflect on why this construction is working.

\section{Rotations, Applications and Insights.}

When we speak about movement between states we speak of rotation. Usually rotation is defined to happen around some axis, but it can also be defined to happen in some plane. It has been shown \cite{HavelDoran} that rotation generated by one of 15 $4\times 4$ two-qubit Pauli matrices can be associated with rotation in one of 15 planes formed by a pair of axes from the set $\{z_1; x_1; y_1; z_2; x_2; y_2 \}$. Let us concentrate on planes formed by axes of the same Bloch Sphere, i.e. local rotations.

To move between the states we apply to an initial state a Unitary of the form $e^{i\frac{\theta}{2}(P)}$ where $P$ is one of 15 two-qubit Pauli matrices and $\theta$ is the angle of rotation. This is equivalent to a rotation of coordinate axes. For example to move from state $\ket{\uparrow\uparrow}$ to the state $\ket{\uparrow\leftarrow}$ we apply $e^{i\frac{\pi}{4}(I \otimes \sigma_y)}$ to the former. This is equivalent to rotation of the coordinate plane formed by the axes $z_2$ and $x_2$ by $\pi/2$ (compare figures \ref{fig:uu} and \ref{fig:ul}). Note how in this representation $\theta$ equals to the angle of rotation on the Bloch Sphere. 

\begin{figure}[H]
  \centering
\begin{minipage}[b][2cm][t]{0.2\textwidth}
\centering
\tdplotsetmaincoords{70}{20}
\begin{tikzpicture}[tdplot_main_coords, scale=0.75]
    % Draw the Bloch sphere with reduced opacity
    \shade[ball color=white!50, opacity=0.6] (0,0,0) circle (1cm);
    
    % Draw axes with adjusted scaling
    \draw[thick,->, scale=1] (0,0,0) -- (1.2,0,0) node[anchor=north, font=\small]{$x_1$};
    \draw[thick,->, scale=1] (0,0,0) -- (0,1.2,0) node[anchor=south, font=\small]{$y_1$};
    \draw[thick,->, scale=1] (0,0,0) -- (0,0,1.2) node[anchor=east, font=\small]{$z_1$};
    
    % Draw equator (dotted line) on the back side with adjusted scaling
    %\tdplotdrawarc[thick, scale=1]{(0,0,0)}{1}{-90}{90}{}{}
    
    % Draw equator (full line) on the front side with adjusted scaling
    \tdplotdrawarc[thin, dashed, scale=1]{(0,0,0)}{1}{-4}{356}{}{}
    
    % Draw the state vector as a thicker arrow with adjusted scaling
    \shade[ball color=orange, opacity=1, scale=2] (0,0,0) circle (0.05cm);
\end{tikzpicture}
\end{minipage}%
\begin{minipage}[b][1.75cm][t]{0.2\textwidth}
\centering
\tdplotsetmaincoords{70}{20}
\begin{tikzpicture}[tdplot_main_coords, scale=0.75]
    % Draw the Bloch sphere with reduced opacity
    \shade[ball color=white!50, opacity=0.6] (0,0,0) circle (1cm);
    
    % Draw axes with inverted directions
    \draw[thick,->, scale=1] (0,0,0) -- (-1.2,0,0) node[anchor=north, font=\small]{$z_2$};
    \draw[thick,->, scale=1] (0,0,0) -- (0,-1.2,0) node[anchor=north, font=\small]{$x_2$};
    \draw[thick,->, scale=1] (0,0,0) -- (0,0,-1.2) node[anchor=west, font=\small]{$y_2$};
    
    % Draw equator (dotted line) on the back side with adjusted scaling
    %\tdplotdrawarc[thick, scale=1]{(0,0,0)}{1}{-90}{90}{}{}
    
    % Draw equator (full line) on the front side with adjusted scaling
    \tdplotdrawarc[thin, dashed, scale=1]{(0,0,0)}{1}{-4}{356}{}{}
    
    % Draw the state vector as a thicker arrow with adjusted scaling
    \shade[ball color=orange, opacity=1, scale=2] (0,0,0) circle (0.05cm);
\end{tikzpicture}
\end{minipage}

  \caption{Two Bloch Sphere representation of the entangled state obtained by cyclic permutation of second BS coordinate axes of the state $\ket{\Psi^-}$ $\{x_2;y_2;z_2\}$ to $\{z_2;x_2;y_2\}$}
  \label{fig:rotPsi-}
\end{figure}
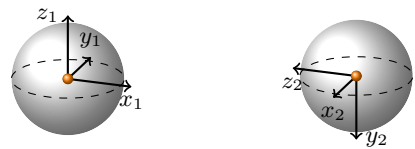

It can be shown how having graphical representation of states and unitaries helps to understand quantum gates, for example CNOT \cite{SFMA2024}. Here we would like to discuss another advantage that requires understanding only of local rotations. Let us look at the Figure \ref{fig:Psi-}. Let's say we are interested in exploring the symmetry of the state $\ket{\Psi^-}$. We want to find some other maximally entangled state that is related to $\ket{\Psi^-}$ in an interesting way. One such state might be the following state: the coordinate axes of the second BS are cyclically permuted (see Figure \ref{fig:rotPsi-}).

From the visual representation we infer that this state is related to $\ket{\Psi^-}$ by $\theta = 2\pi/3$ rotation of the second BS in the plane perpendicular to the axis $z_2 + x_2 + y_2$. The stabilizer group of the state can also be inferred from the representation: $\{I \otimes I; -Z \otimes Y; -Y \otimes X; -X \otimes Z \}$. Guided by equation \ref{eq:rho-group-sum} we can sum the group members to obtain density matrix (eq. \ref{eq:sum}). 

\begin{equation}
    \rho = \frac{1}{4}\begin{pmatrix}
1 & i & -1 & i \\
-i & 1 & i & 1 \\
-1 & -i & 1 & -i \\
-i & 1 & i & 1
\end{pmatrix} 
\label{eq:sum}
\end{equation}

From the density matrix we can obtain the ket form of the state: $\frac{1}{2}\ket{\uparrow\uparrow -i\uparrow\downarrow -\downarrow\uparrow -i\downarrow\downarrow}$.  We can do another cycle of permutations on the second BS and obtain the graphical representation, stabilizer group, density matrix and ket of a state $\frac{1}{2}\ket{\uparrow\uparrow -\uparrow\downarrow -i\downarrow\uparrow -i\downarrow\downarrow}$. Looking only at the ket forms of the two states together with the state $\frac{1}{\sqrt{2}}\ket{\uparrow\downarrow - \downarrow\uparrow}$ one would hardly see that three of them are related by a simple symmetry. 

Next we would like to discuss an observation that might provide a hint to why relations discussed in the paper take place. Let us look at the Figure \ref{fig:rotPsi-} and the stabilizer group of the corresponding state $\{I \otimes I; -Z \otimes Y; -Y \otimes X; -X \otimes Z \}$. 

If we hold the axes $-x_1$ and $z_2$ and rotate the Bloch Spheres simultaneously around those axes in, say, clockwise direction, we keep the relative alignment of all the axes. Remembering that it is the relative alignment of the axes that determines the state, the state hasn't been changed by rotation. Therefore we can say that the stabilizer group indicates the pairs of axes simultaneous rotation about which keeps the state unchanged.

For the case of separable states the interpretation is similar. Let us take the state $\ket{\uparrow\leftarrow}$ with stabilizer group $\{I \otimes I; Z \otimes I; - I \otimes X; - Z \otimes X\}$ depicted in Figure \ref{fig:ul}. $Z \otimes I$ means that rotation about the $z_1$ axis doesn't change the state, same for the $-x_2$ axis. Furthermore the simultaneous rotation about $z_1$ and $-x_2$ axis also leaves the state unchanged. Effectively, the group members of the form $I \otimes P$ link rotation around a given axis with rotation around the statevector arrow, i.e. highlighted direction in space. 

The final observation we make in this work is that, for a separable state, we only know the axes aligned with the statevector direction. For example in Figure \ref{fig:uu} the positions of axes $x_1, y_1, x_2, y_2$ are unknown, and we have arranged them to create a familiar coordinate structure. These axes could point in any direction within the plane orthogonal to the statevector, as long as $x$ and $y$ remain mutually perpendicular. We could say that in the separable mode there is a highlighted direction - direction of statevector arrows - and we have information only about that direction. This observation is consistent with the fact that stabilizer group of a separable state only contains the labels of the axes aligned with the direction of the statevector arrows.

\section{Conclusion}

In this work we have been developing the graphical representation of pure two-qubit states on two Bloch Spheres focusing on maximally entangled stabilizer states. We have shown a way to connect algebraic state form (density matrix) with the graphical representation. This is done via the stabilizer group of the state. Stabilizer group translates in a straightforward way into both algebraic and graphical form. To obtain density matrix from a stabilizer group one just needs to sum the group members and normalize the state. To obtain the graphical representation one just needs to interpret the group members as co-aligned coordinate axes. 

Although in this work the correspondence is explicitly shown only for stabilizer states, it can be extended to arbitrary pure states as any pure state may be expressed through stabilizer states. 

This work brings us closer to two Bloch Sphere visualization that is intuitive and easy to manipulate. It also makes the graphical representations more accessible to quantum computing and quantum foundations communities. The correspondences described in this work allow for exploration of non-trivial symmetries in two-partite Hilbert space and visual analysis of quantum operations. 

The relationship between the stabilizer group and the two Bloch Sphere representation can be understood as, respectively, encoding and depicting the symmetries of rotations that leave the state unchanged.

On the level of quantum foundations the aforementioned correspondence reiterates and solidifies the observation that entanglement is related to inversion of particle's internal coordinates.

\bibliographystyle{apsrev4-2} % Choose bibliography style

\begin{thebibliography}{99} % "99" is just a placeholder for the widest entry label

\bibitem{Bloch1946}
Bloch F., Nuclear Induction, Phys. Rev., vol. 70, pp. 460-474,  \url{https://link.aps.org/doi/10.1103/PhysRev.70.460} (1946)
\bibitem{Feynman1957}
Feynman, R. P., Vernon, F. L., Hellwarth, R. W.; Geometrical Representation of the Schrödinger Equation for Solving Maser Problems. J. Appl. Phys. 1 January 1957; 28 (1): 49–52. \url{https://doi.org/10.1063/1.1722572} (1957)
\bibitem{Radcliffe1971}
Radcliffe J.M., Some Properties of Coherent Spin States, J. Phys. A, vol. 4, pp. 313, (1971)
\bibitem{Arecchi1972}
Arecchi, F. T. et al. Atomic Coherent States in Quantum Optics. Phys. Rev. A 6, 2211 (1972). \url{https://link.aps.org/doi/10.1103/PhysRevA.6.2211}
\bibitem{Bengtsson2006} Bengtsson, I.;  Zyczkowski, K.  Geometry of Quantum States: An Introduction to Quantum Entanglement; Cambridge University Press: Oxford, UK, 2006.
\bibitem{Morelli2024}
Morelli, S., Eltschka, C., Huber, M., Siewert, J. Correlation constraints and the Bloch geometry of two qubits. Phys. Rev. A 109, 012423 (2024). \url{https://doi.org/10.1103/PhysRevA.109.012423}
\bibitem{Mosseri2001}  Rémy Mosseri and Rossen Dandoloff 2001 J. Phys. A: Math. Gen. 34 10243 DOI 10.1088/0305-4470/34/47/324
\bibitem{Bernevig2003} Bogdan A Bernevig and Han-Dong Chen 2003 J. Phys. A: Math. Gen. 36 8325 DOI 10.1088/0305-4470/36/30/309
\bibitem{Mosseri2006} Mosseri, R. (2006). Two-Qubit and Three-Qubit Geometry and Hopf Fibrations. In: Monastyrsky, M.I. (eds) Topology in Condensed Matter. Springer Series in Solid-State Sciences, vol 150. Springer, Berlin, Heidelberg. \url{https://doi.org/10.1007/3-540-31264-1_9}
\bibitem{Uskov2008}
Uskov, D., Rau, A. R. P. Geometric phases and Bloch-sphere constructions for SU(N) groups with a complete description of the SU(4) group. Phys. Rev. A 78, 022331 (2008). \url{https://doi.org/10.1103/PhysRevA.78.022331}
\bibitem{Rau2021}
Rau, A. R. P. Symmetries and Geometries of Qubits, and Their Uses. Symmetry 13(9), 1732 (2021). \url{https://doi.org/10.3390/sym13091732}
\bibitem{Wie2020}
Wie, C-R. Two-Qubit Bloch Sphere. Physics 2(3), 383-396 (2020). \url{https://doi.org/10.3390/physics2030021}
\bibitem{Majorana1932} Majorana, E. Atomi orientati in campo magnetico variabile. Nuovo Cim 9, 43–50 (1932). https://doi.org/10.1007/BF02960953
\bibitem{Aulbach2010} Martin Aulbach et al 2010 New J. Phys. 12 073025 DOI 10.1088/1367-2630/12/7/073025
\bibitem{Cormann2017} Mirko Cormann and Yves Caudano 2017 J. Phys. A: Math. Theor. 50 305302 DOI 10.1088/1751-8121/aa7639
\bibitem{Yu-Guo2021} Yu-Guo Su et al 2021 Chinese Phys. B 30 030303 DOI 10.1088/1674-1056/abc2bc
\bibitem{Xingyu2021} Xingyu Zhang et al 2022 Commun. Theor. Phys. 74 065102 DOI 10.1088/1572-9494/ac6801
\bibitem{SFMA2024}
Filatov S, Auzinsh M. Towards Two Bloch Sphere Representation of Pure Two-Qubit States and Unitaries. Entropy. 26(4):280. (2024) https://doi.org/10.3390/e26040280 
\bibitem{NielsenChuang2011}
Michael A. Nielsen and Isaac L. Chuang, Quantum Computation and Quantum Information: 10th Anniversary Edition, Cambridge University Press (2011), ISBN: 9781107002173
\bibitem{Gottesman1997}
D. Gottesman, "Stabilizer codes and quantum error correction," quant-ph/9705052, Caltech Ph.D. thesis (1997).
\bibitem{Gross2006}
Gross, D.; Hudson’s theorem for finite-dimensional quantum systems. J. Math. Phys. 1 December 2006; 47 (12): 122107. https://doi.org/10.1063/1.2393152
\bibitem{Garcia2014}
Hector J. Garcia, Igor L. Markov, and Andrew W. Cross, "On the geometry of stabilizer states", \textit{Quantum Information and Computation}, Vol. 14, No. 7-8, pp. 683-720 (2014). doi: \url{https://doi.org/10.26421/QIC14.7-8-9}
\bibitem{Wu2015} Wu, Xia, et al. "Determination of stabilizer states." Phys. Rev. A, vol. 92, no. 1, pp. 012305, Jul 2015. DOI: 10.1103/PhysRevA.92.012305
\bibitem{HavelDoran} Havel, T.F.; Doran, C.J.L. A Bloch-sphere-type model for two qubits in the geometric algebra of a 6D Euclidean vector space. In Quantum Information and Computation II; SPIE: Bellingham, WA, USA, 2004; Volume 5436. https://doi.org/10.1117/12.540929
% Add more references as needed, following the same format

\end{thebibliography}

\end{document}